\pdfoutput=1
\documentclass[11pt,a4paper]{article}
\usepackage{swisstext2021}
\usepackage{times}
\usepackage{latexsym}

\usepackage{lipsum}
\usepackage{graphicx}
\usepackage[utf8]{inputenc}
\usepackage{mathptmx}
\usepackage{hyperref}
\usepackage[scaled=.90]{helvet}
\usepackage{courier}
\usepackage{color, colortbl}
\usepackage{booktabs}
\usepackage[T1]{fontenc}
\usepackage{makecell}
\usepackage{multirow}
\usepackage{subcaption}
\usepackage{tikz}
\usepackage{pgfplots}
\pgfplotsset{compat = 1.9}
\pgfplotsset{major grid style={dashed}}
\usepackage{siunitx, mhchem}
\usepackage{amsmath}
\usepackage{url}

\newcommand{\DEVELOPMENT}{1} 
\usepackage{ifthen}
\ifthenelse{\DEVELOPMENT = 1}{
	\newcommand{\aga}[1]{\textcolor{purple}{\textbf{AA:} #1}}	
    
    \newcommand{\ya}[1]{\textcolor{cyan}{\textbf{YA:} #1}}
    \newcommand{\ok}[1]{\textcolor{red}{\textbf{OK:} #1}}
}{
	\newcommand{\aga}[1]{}
	\newcommand{\sg}[1]{}
	\newcommand{\ya}[1]{}
	\newcommand{\ok}[1]{}
}

\usepackage{microtype}

\aclfinalcopy

\title{Dialectal Speech Recognition and Translation of \\
Swiss German Speech to Standard German Text:\\
Microsoft's Submission to SwissText 2021}

\author{Yuriy Arabskyy,  Aashish Agarwal, Subhadeep Dey,  Oscar Koller \\
  Microsoft - Munich, Germany \\
  \texttt{\{yuarabsk, t-aagarwal, subde, oskoller\} @microsoft.com} \\}
\date{}

\begin{document}
\maketitle
\begin{abstract}
This paper describes the winning approach in the Shared Task 3 at SwissText 2021 on Swiss German Speech to Standard German Text, a public competition on dialect recognition and translation.
Swiss German refers to the multitude of Alemannic dialects spoken in the German-speaking parts of Switzerland. Swiss German differs significantly from standard German in pronunciation, word inventory and grammar. It is mostly incomprehensible to native German speakers. Moreover, it lacks a standardized written script.
To solve the challenging task, we propose a hybrid automatic speech recognition system with a lexicon that incorporates translations, a 1\textsuperscript{st} pass language model that deals with Swiss German particularities, a transfer-learned acoustic model and a strong neural language model for 2\textsuperscript{nd} pass rescoring.
Our submission 
reaches 46.04\% BLEU on a blind conversational test set and outperforms the second best competitor by a 12\% relative margin.
\end{abstract}

%
%
\section{Introduction}
%
%

While general speech recognition has matured to a point where it surpasses human performance on specific datasets \cite{xiong2016achieving}, dialectal recognition
as in the case of Swiss German \cite{nigmatulina-etal-2020-asr} or Arabic dialects \cite{ali_connecting_2021,hussein_arabic_2021,ali_multidialect_2018} still represents a major challenge.
Swiss German refers to the multitude of Alemannic dialects spoken in the German-speaking parts of Switzerland. It hence represents dialects that differ significantly from standard or high German in pronunciation, word inventory and grammar. Moreover, it lacks a standardized writing system. High German is commonly used for the large majority of written communication between people and in the media of German-speaking Switzerland, while in rather informal chats and short messaging Swiss people may use transliterated non-standardized Swiss dialect.
The process of transcribing Swiss German into high German text therefore requires speech recognition with an inherent translation step. Moreover, the task can be considered low-resource as available data remains extremely scarce. 

In previous studies, \newcite{garner_automatic_2014} tackled this challenge by training hybrid models (HMM-GMM, HMM-DNN, and KL-HMM) to transcribe Walliserdeutsch, a Swiss-German dialect spoken in the south-western alpine canton of Switzerland and further used a phrase-based machine translation model to translate it to standard German.
Following this, other researchers explored techniques to add the translation step in the lexicon by directly mapping Swiss-German pronunciation to standard German. 
\newcite{stadtschnitzer2018adaptation} estimated Swiss German pronunciations from a standard German speech recognition model using a data-driven technique, and trained stronger TDNN-LSTM based acoustic models.
\newcite{kew2020uzh} and \newcite{nigmatulina-etal-2020-asr} trained transformer-based G2P models from standard German to Swiss pronunciations and trained a Kaldi-based TDNN+ivector system using the WSJ recipe\footnote{\url{https://github.com/kaldi-asr/kaldi/tree/master/egs/wsj}}.
Yet a third approach is to directly apply end-to-end deep learning models. \newcite{buchi2020zhaw} and \newcite{agarwal2020ltl} at SwissText 2020 used the Jasper architecture \cite{li2019jasper} and Mozilla DeepSpeech \cite{hannun2014deep}, respectively.
In both cases the system was first trained on high German data and then transfer-learned to Swiss German.

In this paper, we describe our proposal to solve the challenging task of transcribing Swiss German speech to standard German text. It won the competition at SwissText 2021 with a large margin to other competing systems.

%
%
\section{System Overview}
%
%

In this section, we propose our changes to a conventional hybrid~\cite{bourlard_new_1996} automatic speech recognition (ASR) system, which relies on lexicon and alignments for good performance. We present details in order to enable it for dialect speech recognition and translation.

%
%
\subsection{Data}\label{sec:data}
%
%

To train our proposed model, we utilized a selection of publicly available and internal datasets.
Our starting point was the Swiss Parliament Corpus V2 dataset \cite{pluss2020swiss} shared as part of the SwissText 2021 competition. It covers 293 hours and contains recordings from the local parliament of the Kanton Bern. Its transcripts are in standard German while the audio covers Swiss German (predominantly in the Bernese dialect). The dataset has been preprocessed by the publishers with the purpose of cleaning its annotations and ensuring a good match between audio content and transcription. It is provided with a choice of different preprocessing flavors.
We used the train\_all split. 
In addition, we used a 493-hour internal dataset representing a media domain encompassing conversational speech from interviews, discussions, podcasts and others.
A subset (around 50 hours) of the data is annotated with both Swiss transliterations as well as standard German. The remaining data has only been annotated with standard German.
Additionally, we used an internal high German dataset encompassing around 10k hours to pre-train our model.

 In terms of test data, the SwissText 2021 competition was accompanied by a 13 hours conversational test set covering Swiss German speakers from all German-speaking parts of Switzerland. The encountered dialectal distribution is claimed to closely match the real distribution in Switzerland. The set was not disclosed to the participants. 
 Hence, for the analysis in this paper, we
 report our numbers on a publicly available test set part of the dataset from the Bernese parliament~\cite{pluss2020swiss}. It comprises 6 hours of dialectal speech.

%
%
\subsection{Lexicon}
%
%

\begin{table*}
    \centering
    \small
    \begin{tabular}{cccccc}
    \toprule
         \bf 2nd person plural & \bf 2nd person sing & \bf Diminuation & \bf Shortening & \bf Translation & \bf Variability  \\
         \midrule
              fragt &
              fragst &
              erdmännchen &
              gymnasium & 
              kopf &
              kannst \\
              
               \emph{f hr a\_ g ax t} &
               \emph{f hr a\_ k sh} &
               \emph{e\_r t m eh n l i\_} &
               \emph{g ih m i\_} &
               \emph{g hr ih n t} &
               \emph{k a sh} \\
              
              \addlinespace[2mm]
              riecht &
              riechst &
              gläschen &
              schwimmbad &
              kneipe &
              zweites \\
              
               \emph{sh m oe k c ax t} &
               \emph{sh m oe k sh} &
               \emph{g l e\_ s l i\_} &
               \emph{b a\_ d ih} &
               \emph{b ai ts} &
               \emph{ts v ai t} \\
         \bottomrule
    \end{tabular}
    \caption{Example words and pronunciations from each G2P test condition}
    \label{tab:g2p_test_data}
\end{table*}

We propose to incorporate the translation from Swiss German to standard German as part of the lexicon. 
However, this leads to a complex and often ambiguous mapping from phoneme to grapheme sequences, which is very different from languages with a direct relation between writing scheme and pronunciation (e.g. English or standard German). Subsequently, statistical models that map graphemes to phonemes (G2P) trained on Swiss German data incorporating such translations yield much noisier output with significantly higher phone error rates as compared to G2Ps for standard languages. 
To mitigate this problem, we construct the lexicon in several stages. 

In a first step, we make use of parallel corpora encompassing Swiss and standard German annotations to extract word mappings between Swiss and standard German. 
Sophisticated filtering methods help to ensure a high quality of these mappings. We opt for frequency filtering and filtering based on vicinity in a word embedding space~\cite{bojanowski2017enriching} of Swiss German words taking the most frequent mapping as center point. 

In a second step, a standard German G2P model is applied to convert Swiss German transliterations into corresponding phone sequences. This results in a dictionary that maps standard German words to Swiss pronunciations. 
Jointly with existing Swiss German lexicon resources~\cite{schmidt_swiss_2020}, the previously generated mappings are then used to 
train a dedicated Swiss German G2P model.

We evaluate the quality of the resulting G2P model on a manually labeled test set. Those cover mappings from standard German words to Swiss German phone sequences and encompass a variety of relevant categories such as diminuation, shortening or translation.
Refer to Table \ref{tab:g2p_test_data} for samples of the assessed categories. 

The Swiss G2P model allows to find suitable pronunciations for 
the relevant word inventory present in the acoustic and language model training corpora. However, to further increase the quality of the given pronunciations, data-driven lexicon learning techniques \cite{zhang_acoustic_2017} are applied. Those help
to identify and correct noisy lexicon entries.

%
%
\subsection{Language Model}\label{sec:lm}
%
%

Incorporating the translation from Swiss German to standard German as part of the lexicon introduces significant ambiguity in the decoding process. To counteract, we suggest using a strong standard German language model (LM) which helps to produce accurate hypotheses. 
We employ a first pass count-based LM to output up to 100 sentence hypotheses and a second pass neural LSTM (long short-term memory) LM~\cite{sundermeyer_lstm_2012} for rescoring~\cite{deoras_fast_2011}.
The first pass model is a 5-gram LM trained
on large amounts of standard German text corpora totalling to over 100 billion words. We apply Kneser-Ney smoothing \cite{kneser_improved_1995}. 

Furthermore, we make some adjustments to better deal with Swiss German particularities, as described in the following paragraphs.

\paragraph{Compounds:} German is a compounding language and tends to compose words (particularly nouns) of several smaller subwords. The resulting chains of word stems 
can lead to an infinitely large vocabulary size with words that occur very infrequently throughout the corpus. This spreading of probability mass weakens the LM. We hence decompound all compounded words in the training corpus and split them into subwords.

\paragraph{Clitics:} Swiss German tends to merge words beyond compounding, not preserving word stems~\cite{hollenstein_compilation_2014}. For instance, the Swiss German `\emph{hemmer}' 
is the translation of `\emph{haben wir}' in standard German (English: `\emph{have we}'). We identified approximately $8000$ clitics in our corpus. We incorporate them in the decoding process by updating lexicon and LM. Following the example above the translated clitic `\emph{haben\#wir}` with the corresponding Swiss pronunciation is added to the lexicon.
As for the LM, we merge occurrences of relevant word pairs and interpolate with the unmerged LM.

%
%
\subsection{Acoustic model}
%
%

The acoustic model is trained with 80 dimensional log-mel filterbank features, with a processing window of 25ms and 10ms frame shift. The feature vector from the previous frame is concatenated with the current frame to obtain a 160 dimensional vector.
We used a LC-BLSTM (latency controlled bidirectional long short-term memory) based acoustic model, that is popularly applied in speech recognition for controlling decoding latency to a few frames~\cite{chen7}. The model was trained with alignments from a feed-forward network with context-dependent tied states (senones). The model has $\sim$9k senone units. The LC-BLSTM is trained with 6 hidden layers with 512 units each. The hidden vectors from the forward and backward propagation were concatenated and then projected to a 512 dimensional vector. The model is trained with a cross entropy loss function.
The decoding lexicon is extended with Swiss German words for training. The transliterations are used during forced alignment
whenever possible. This helps to reduce the pronunciation ambiguity in the alignment phase and is especially helpful in the early
training phases when no strong model is available for alignment.

The results are reported in terms of BLEU \cite{papineni_bleu_2002} and word error rate (WER) on the Swiss Parliament test set described in Section \ref{sec:data}.

%
%
\section{Results and Discussions}
%
%

 \begin{table*}[ht]
     \centering
     \small
     \begin{tabular}{c l cc}
     \toprule
        \bf Row & \bf Description & \bf  \bf WER & \bf BLEU \\
          \midrule
          1 &  Swiss Parliament & 45.93 &  33.16 \\
          2 & ~~~+ Transfer Learning  & 42.09 & 37.53 \\
          3 & ~~~~~~+ Internal Data & 41.10 & 38.49  \\
          4 &  ~~~~~~~~~+ 2\textsuperscript{nd} Pass Rescoring & 38.71 & 41.91  \\  
          \bottomrule
     \end{tabular}
     \caption{Performance in [\%] of different system configurations evaluated on the Swiss Parliament test set.}
    \label{tab:results_chparl}
 \end{table*}

An ablation study of the proposed approaches is presented in Table~\ref{tab:results_chparl}. All of the performance gains
in this section will be reported as relative percentage improvements, while aforementioned table contains absolute numbers.

We first evaluate the effect of transfer learning on the results with the Swiss Parliament training set.
It can be observed that it significantly helps to improve both WER and BLEU. In particular, the transfer-learned model (row 2, Table~\ref{tab:results_chparl}) improves over the model trained from scratch (row 1, Table~\ref{tab:results_chparl}) by $8.4\%$ WER and $11.6\%$ BLEU. The result shows that a well-trained German model can effectively boost the limited resources of Swiss German.

Further adding additional internal training data shows additional gains in performance. As such, we observe that the WER improves by $2.3\%$ and BLEU by $2.5\%$.

Finally, 2\textsuperscript{nd} Pass rescoring is applied as described in Section~\ref{sec:lm} to reorder the top 100 hypotheses.
It can be observed from row 4, Table~\ref{tab:results_chparl} that rescoring helps to improve the performance
by $5.8\%$ WER and $8.9\%$ BLEU.

Our submission to SwissText 2021 achieves $46.04\%$ BLEU on the
official SwissText blind test set. 
This leads to a $12\%$ relative margin in BLEU with respect to the second best competitor which was $40.99\%$. 

The acoustic models have been trained using 8 GPUs for 25 epochs. This results in a total training time of around 400 GPU-hours when training on Swiss Parliament only and about 1200 GPU-hours when adding the internal data.

%
%
\section{Conclusion and Future Work}
%
%

In this paper, we described a speech recognition system that achieves strong results on the task of recognizing Swiss German dialect and translating it into standard German text. 
We proposed a hybrid ASR system with a lexicon that incorporates translations, a 1\textsuperscript{st} pass language model that deals with Swiss German word compounding and clitics, an acoustic model that is transfer-learned from standard German resources and a strong neural language model for 2\textsuperscript{nd} pass rescoring to smoothen translation artifacts.  
Furthermore, we provided an ablation study that allows to infer the effect of adding training data, performing transfer learning and 2\textsuperscript{nd} pass rescoring.
Our submission
reached 46.04\% BLEU on a challenging conversational test set and outperformed all competing approaches by a large margin.

In terms of future work, we would like to investigate word re-orderings as part of the translation, which our current model does not actively support. For instance, Swiss German frequently moves verbs in relative clauses to different positions with respect to the standard German word order.
Furthermore, sequence discriminative training is a promising route for exploration as well as using unsupervised data for acoustic model training.

\bibliography{anthology,acl2020,koller}
\bibliographystyle{acl_natbib}

\end{document}